# Interpretations of ground-state symmetry breaking and strong correlation in wavefunction and density functional theories

John P. Perdew[1,2,*], Adrienn Ruzsinszky[1], Jianwei Sun[3], Niraj K. Nepal[1], and Aaron D. Kaplan[1,#]

[1] Department of Physics, Temple University. Philadelphia, PA 19122.

[2] Department of Chemistry, Temple University, Philadelphia, PA 19122.

[3] Department of Physics, Tulane University, New Orleans, LA 70118.

* perdew@temple.edu

# kaplan@temple.edu

Strong correlations within a symmetry-unbroken ground-state wavefunction can show up in approximate density functional theory as symmetry-broken spin-densities or total densities, which are sometimes observable. They can arise from soft modes of fluctuations (sometimes collective excitations) such as spin-density or charge-density waves at non-zero wavevector. In this sense, an approximate density functional for exchange and correlation that breaks symmetry can be more revealing (albeit less accurate) than an exact functional that does not. The examples discussed here include the stretched $H_2$ molecule, antiferromagnetic solids, and the static charge-density wave/Wigner crystal phase of a low-density jellium. It is shown that (and in what sense) the static charge density wave is a soft plasmon.

This article presents examples and interpretations of broken symmetry and strong correlation, complexities that sometimes show up in quantum mechanical systems of many particles, including molecules and solids. Figure 1 suggests why several different perspectives can be helpful to understand complex phenomena.

The density functional theory (DFT) of Kohn and Sham [1] is an exact-in-principle mean-field-like theory for the ground-state (lowest) energy and electron density (or spin densities) of any $N$-electron system with a Hamiltonian or energy operator of the form

$$\hat{H} = \sum_{i=1}^{N}[-\nabla_i^2/2 + v(\boldsymbol{r}_i)] + \sum_i \sum_{j>i}|\boldsymbol{r}_i - \boldsymbol{r}_j|^{-1} \quad (1)$$

(taking fundamental constants to be 1). $v(\boldsymbol{r})$ is an external scalar potential, such as the Coulomb attraction to the nuclei, and the electron-electron Coulomb repulsion is explicitly included. A ground-state $N$-electron wavefunction is an eigenstate of the Hamiltonian for the lowest energy eigenvalue. Since electrons are spin-1/2 fermions, the physical wavefunctions must be antisymmetric, changing sign under the exchange of any two electrons. Squaring a wavefunction yields a positive probability distribution. Integrating the square of the wavefunction over $N$-1 electron position vectors and $N$-1 spin coordinates yields a one-electron spin density $n_\sigma(\boldsymbol{r})$ (where $\sigma = \uparrow$ or $\downarrow$ along the $z$-axis of spin quantization) and total electron density $n(\boldsymbol{r}) = n_\uparrow(\boldsymbol{r}) + n_\downarrow(\boldsymbol{r})$ that define the starting point for DFT. Integrating over all but two electron positions $\boldsymbol{r}$ and spins $\sigma$ yields an electron pair density $\rho_{\sigma\sigma'}(\boldsymbol{r},\boldsymbol{r}')$ that shows explicitly how the electrons avoid one another due to wavefunction antisymmetry (which makes $\rho_{\sigma\sigma}(\boldsymbol{r},\boldsymbol{r}) = 0$) and Coulomb repulsion. Although the Hamiltonian of Eq. (1) does not depend explicitly on spin, the spin-state importantly affects the ground-state energy through the antisymmetry of the wavefunction.

There is an exact ground-state density functional for the energy, including non-interacting kinetic,

external, Hartree, and exchange-correlation terms, whose minimization leads to exact one-electron Schrödinger equations for the orbitals or one-electron wavefunctions used to construct the non-interacting kinetic energy and the electron density. A practically useful method is attained by approximating only the negative exchange-correlation energy functional $E_{xc}[n_\uparrow, n_\downarrow]$, sometimes based on exact constraints and on appropriate norms such as the uniform electron gas for which all functionals discussed here are exact.

Coulomb correlation means all effects beyond the symmetry-unbroken self-consistent mean-field Hartree-Fock approximation [2]. The mean-field appearance of Kohn-Sham theory too often creates the impression that this theory lacks correlation or at least strong correlation. But Coulomb correlation is about as strong as exchange even in ordinary *sp*-bonded molecules, and this correlation is well described by non-empirical approximate functionals [1,4-7], as Table 1 shows. Kohn-Sham DFT achieves computational efficiency by *not* calculating a correlated many-electron wavefunction, but it still includes the Coulomb-repulsion-driven correlation among electrons in its functional or rule for finding the exchange-correlation term of the total energy from the electron up- and down-spin densities, and in its related exchange-correlation potential in the one-electron Schrödinger equation that shapes the orbitals and the spin densities in a self-consistent calculation.

"Strong correlation" is sometimes used to mean "everything that density functional theory gets wrong". Yet hybrid functionals (including part of the Hartree-Fock exchange) like HSE06 [9,10] (for non-metallic states) and meta-GGA's like SCAN [6,11-17] are yielding quantitatively correct ground-state (and we emphasize "ground-state") results for some systems that have long been regarded as strongly correlated. In the cuprate high-temperature superconducting materials, for example, the SCAN meta-GGA [6] is able to do what simpler density functionals (LSDA, GGA) cannot [13], by creating the correct spin moments on the copper atoms, their antiferromagnetic order, and a correctly non-zero band gap in the undoped material that correctly disappears under the doping that also leads to superconductivity [12-14]. Full but more reliable self-interaction corrections [7] might eventually further improve approximate DFT for other strongly-correlated systems. And, as will be argued here, even the simplest DFT approximations yield a qualitative insight into some such systems.

The first interpretation to be expressed here is that certain "strong correlations" that are present as fluctuations in the exact symmetry-unbroken ground-state wavefunction are "frozen" in symmetry-broken electron densities or spin-densities of approximate DFT. For finite systems, this would not happen with the exact functional. So, while the exact functional would always be exact for the ground-state energy and density, it would not always be as qualitatively revealing as the approximate one.

This first interpretation (but without the term "strong correlations") is eloquently expressed in P.W. Anderson's essay "More is Different" [18], which explains that, while a ground-state wavefunction is necessarily static, it must describe fluctuations in the expectation values of operators that are not diagonalized along with the Hamiltonian, and that these fluctuations can freeze as the number of particles in the system grows large, leading to an observable symmetry breaking that would not be possible in a few-particle system. Symmetry-breaking in approximate DFT can thus be more correct for solids than it is for small molecules, but in either case it can reveal a strong correlation in a symmetry-unbroken wavefunction. The possibility of observable symmetry breaking is compatible [19] with the continued existence of a symmetry-unbroken exact ground-state wavefunction. Symmetry breaking is surprising only in quantum physics; in classical physics, it is familiar and intuitive.

The second interpretation involves the wave-like fluctuations and collective excitations of a solid. A wave in the total electron density $n(\boldsymbol{r}) = n_\uparrow(\boldsymbol{r}) + n_\downarrow(\boldsymbol{r})$ is a plasma wave (quantized as plasmons), and a wave in the net electron spin density $m(\boldsymbol{r}) = n_\uparrow(\boldsymbol{r}) - n_\downarrow(\boldsymbol{r})$ is a spin density wave. Like other waves, these have amplitude, wavelength $\lambda$ or wavevector q = $2\pi/\lambda$, and frequency $\omega$, and at small amplitudes can be simply superposed or added together. However, their frequencies are not sharply defined. By the uncertainty principle, a smaller frequency width leads to a longer lifetime. Collective excitation modes are distinguished from other fluctuation modes by much smaller frequency widths. The second interpretation explains how some symmetry breakings and strong correlations can arise: Under a variation of the external potential, a collective excitation or fluctuation of the electrons of non-zero wavevector, such as a charge-density wave or a spin-density wave, can soften, with an excitation energy or frequency tending to zero, until it appears as a static wave in the symmetry-broken density or spin-density of an approximate functional. Since the frequency is positive, the frequency width must go to zero when the frequency does. This mechanism is analogous to the softening of a phonon mode of non-zero wavevector that can lead to a distortion or structural phase transition of an ionic lattice. The ground-state energy of a quantum lattice nearly equals that of the classical lattice plus the zero-point energy of its lattice vibrational modes. Via the fluctuation-dissipation theorem [20,21], the ground-state total energy of a system of interacting electrons, including the Coulomb correlation energy, has contributions from fluctuations of various wavevectors, including the non-negative zero-point energies $\hbar\omega/2$ of its collective excitations. Collective excitations, like observable or physical symmetry breakings, are emergent phenomena, arising only in systems with large numbers of particles [22].

The Hamiltonian of Eq. (1) can break spatial symmetries through its external potential, but that is not the subject of this discussion. We also note that this Hamiltonian, while a good starting point, is not exact. It omits relativistic effects including the spin-orbit interaction, and it neglects the fact that the external potential comes from nuclei that can move and, in a solid, can break lattice symmetries. These physical effects will be ignored here for the sake of clarity and simplicity.

We know that it is possible to find the eigenstates of a complete set of commuting observables including the Hamiltonian. The states with the lowest eigenvalue of the Hamiltonian are the ground states. If the ground state is non-degenerate, the ground-state density will have all the symmetries of the external potential. For example, the ground-state of the Ne atom is non-degenerate, and its electron density has the spherical symmetry of the $v(r) = -10/r$ Coulomb attraction to the nucleus. If the ground state is degenerate, an equally-weighted average (equi-ensemble) of all ground-state densities or spin densities will have the symmetries of the external potential. "The symmetry of the Hamiltonian is the symmetry of the ground state" [19].

The Hamiltonian of Eq. (1) is independent of electron spin. Then the complete set of commuting observables can include the square and $z$-component of the total electron spin. When the ground-state is non-degenerate, as in a closed-shell atom or molecule, or presumably in a finite non-ferromagnetic and non-ferrimagnetic crystal, the square of the total spin must be zero ($S = 0$ or spin singlet state) and thus the local net spin density must be zero, and the exact ground-state density is expected to have all the symmetries of the external potential. When there is a net spin moment or a net current density, of course, the ground-state must be degenerate, and a single ground state from a degenerate set need not have the symmetry of the external potential, although the state of thermal equilibrium in the zero-temperature limit still should have it.

A closed-shell molecule like $H_2$ has a non-degenerate ground-state whose exact electron density must remain spin-unpolarized at all bond lengths. In 1976, Gunnarsson and Lundqvist [23] made an LSDA calculation for the ground-state $H_2$ molecule at various bond lengths. For bond lengths less than a critical value (1.7Å), they found a spin-unpolarized density and a realistic binding energy curve. Above the critical bond length, they found an unobservable spin symmetry breaking, with the net spin up near one nuclear center and down near the other (a kind of molecular precursor of the antiferromagnetism to be discussed in the next section). The whole binding-energy curve was realistic only when the symmetry of the spin density was allowed to break under bond stretching. A constraint that prevented spin symmetry breaking led in the limit of infinite bond length to an LSDA total energy much higher than the LSDA energy of two individual hydrogen atoms.

The exact wavefunction tells us that, at a stretched but finite bond length, when an electron with a given spin direction is near one nucleus, an electron with the opposite spin direction is highly likely to be near the other. This is a strong correlation within a symmetry-unbroken wavefunction of zero total spin, which is revealed to us by the breaking of the symmetry of the spin density in the LSDA calculation.

Similarly, symmetry suggests that the exact ground-state density of any finite system with zero net spin moment should be spin-unpolarized. But approximate spin-density functional calculations, e.g., [11-15,17], find that many materials and especially transition-metal oxides are antiferromagnetic, with alternating net spin moments on alternating atomic sites. This is presumably a strong correlation within the exact symmetry-unbroken ground-state wavefunction that freezes into the symmetry-broken ground-state spin densities of the approximate density functional. Strong correlations within an exact symmetry-unbroken wavefunction might tell us that there are local spin moments on the transition-metal atoms that are correlated, so that for example a spin-up moment on a given atom has spin-down neighbors on the nearest-neighbor atoms, although the direction of the spin moment on a given atom is not fixed.

A connection between spin-density waves and antiferromagnetism was proposed by Overhauser [24]. Our interpretation is that a spin-density wave drops down (as a function of lattice constant) in energy and frequency to create the broken-symmetry static spin density. Under extreme compression of a given lattice, any material will be a non-magnetic metal with zero energy gap and no local magnetic moment. As this compression is relaxed, we imagine that a spin density wave will soften. To create an ordered antiferromagnetic state, the non-zero wavevector of the soft spin density wave must extend from the center of the first Brillouin zone of the underlying ionic lattice to the center of one of the Brillouin-zone faces, typically doubling the size of the unit cell in real space and opening an energy gap that makes the material an insulator. In approximate Kohn-Sham density functional theory, the metallic, non-magnetic state of unbroken symmetry remains as a solution of the Kohn-Sham equations, but not as the solution of lowest energy. Many correlations become stronger and symmetries tend to break as the electron density and electron kinetic energy decrease.

Anderson [25] found an accurate estimate of the ground-state energy for an antiferromagnetic Heisenberg model Hamiltonian describing the effective interaction between localized spin moments on a simple cubic lattice. He started from the energy of a classical antiferromagnetic ground state, then included the zero-point energy of the transverse spin-wave excitations. Here the transverse directions are perpendicular to the $z$ direction, i.e., to the spin direction on the first of the two spin sub-lattices. The frequency is of course zero for any long-wavelength mode (Goldstone mode) that simply rotates the spin direction, since there is no restoring force. The expectation value of the spin moment vector on each site of a sublattice can point in any direction, as long as the expectation value on each site of the other sublattice points in the opposite direction. Anderson estimated the time (proportional to the number of atoms present) for the spin direction to rotate by 90 degrees as three years, much longer than the time scale of a neutron diffraction experiment. Thus the singlet spin symmetry is well broken by the soft spin density wave of non-zero wavevector, and the spin-isotropy symmetry is well broken by the soft Goldstone mode of zero wavevector.

Interestingly, the local spin moment on a transition-metal atom can persist even above the magnetic disordering temperature, both in ferromagnetic iron [26] and in transition metal monoxides and other strongly-correlated materials (where it can still give rise to an insulating gap in the Kohn-Sham density of one-electron states) [11-15].

Now consider a standard model for a simple metal, an infinite jellium in its ground state. The positive background charge is uniform and rigid, and there should always be a ground-state wavefunction with a uniform electron density. But, at low density ($r_s \approx 69$, where the density is $n = 3/[4\pi r_s^3] = k_F^3 / 3\pi^2$ and thus $r_s$ is the radius of a sphere containing on average one electron), where the jellium is greatly expanded from its equilibrium density ($r_s \approx 4$), an exchange-correlation kernel-corrected linear-response density functional calculation [27] finds a static charge-density wave. Approximate density functionals like LSDA also predict [28] a charge-density wave instability of the uniform electron density, but often at the wrong density. The charge-density wave is an incipient Wigner lattice, in the sense that it has a wavevector $q \approx 2.28 k_F$, close to the first reciprocal lattice vector of a bcc lattice, so that a bcc Wigner lattice can be constructed by superposing charge density waves with wavevectors along the twelve (110) directions. At the density where the Wigner lattice first appears ($r_s \approx 85 \pm 20$ from a quantum Monte Carlo calculation [29]), each electron is spread out over its whole Wigner-Seitz cell, but in the $r_s \to \infty$ limit, each electron further localizes to the center of its cell. The Wigner lattice is a strong correlation within the exact symmetry-unbroken ground-state wavefunction at low density, and is justified by the theory of strictly-correlated electrons [30].

The charge density wave in jellium is a case in which we can computationally identify the collective mode or fluctuation and see how its excitation energy or frequency approaches zero as the external potential is varied. The charge-density wave appears [27] to arise from a soft plasmon: a collective density-wave excitation of the system that drops down in excitation energy or frequency and eventually freezes in the broken-symmetry density of an approximate density functional calculation. The argument for this in Ref. [27] was based on extrapolation of the plasmon frequency $\omega_p(q)$ into the range of wavevectors $q$ where the plasmon frequency is mixed with the continuum of single particle-hole excitations, and the plasmon is no longer a true collective excitation. Figures 2 and 3, however, show that this extrapolation is well justified. Figure 2 shows the average frequency of a density fluctuation of wavevector $q$, and Fig. 3 shows the root mean square deviation from this average, both computed from the structure factor [20,21] or spectral function [22] $S(q,\omega)$ of time-dependent density functional theory [22].

The refinement of approximate density functionals toward the exact functional offers the possibility of suppressing symmetry breaking [31], but the result would not be an unmixed blessing since qualitative insights into strong correlation could be lost.

We can now propose a definition of "strong correlation": Strong correlation in a symmetry-constrained wavefunction is any correlation between electrons that results in an exceptionally structured electron pair density, or is otherwise qualitatively different from the "normal" Coulomb correlation found in simple $sp$-bonded materials in their ground states near equilibrium nuclear geometries and having Hamiltonians of the form of Eq. (1). This definition has little in common with "everything that DFT gets wrong". Indeed, DFT with standard approximate functionals perfectly captures the energetic consequences of strict correlation [30] in a uniform electron gas of very low density. Strong correlation by this definition is sometimes displayed by symmetry breaking in a symmetry-unconstrained wavefunction or density. DFT increasingly describes the ground-state energies and densities of systems with $d$ or $f$ electrons that are widely regarded as strongly-correlated.

There is another valid interpretation [32] of spin symmetry breaking in spin-density functional theory: Approximate spin density functionals reliably predict the total electron density, but predict the on-top electron pair density $\sum_{\sigma,\sigma'} \rho_{\sigma,\sigma'}(\boldsymbol{r},\boldsymbol{r})$ (yielding the probability to find two electrons together at the same position in space) more reliably than they predict the net spin density; such functionals predict both reliably when the correlation is not strong. This interpretation is now being used to merge the wavefunction and density-functional approaches [33].

**Acknowledgments:**

JPP and JS acknowledge stimulating discussions of strong correlation with Alex Zunger.

**Funding:**

The work of JPP was supported by the U.S. National Science Foundation under grant number DMR-1939528. The work of AR and NKN was supported by the U.S. National Science Foundation under grant number DMR-1553022. The work of JS was supported by the Department of Energy, Office of Science, Basic Energy Sciences grant number DE-SC0019350. The work of ADK was supported by the Department of Energy, Basic Energy Sciences, through the Energy Frontier Research Center for Complex Materials from First Principles, grant number DE-SC0012575.

**Author contributions:**

J.P.P., A.R., and J.S. designed the research, performed the research, and wrote the manuscript. N.K.N. and A.D.K. designed some of the research and performed the calculations and analysis.

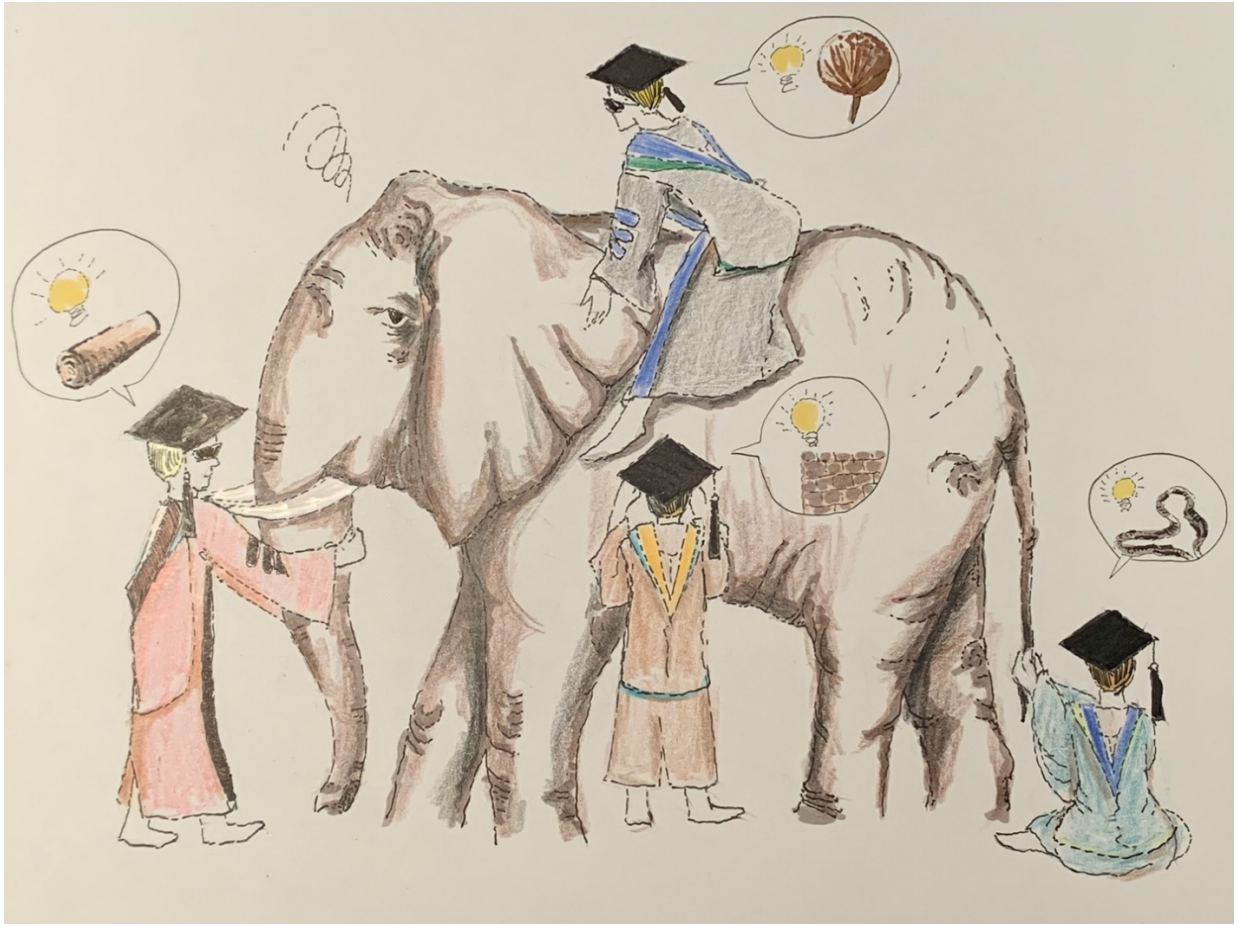

**Fig. 1.** The parable of the blind men and the elephant suggests that scientists who study the same problem from different perspectives should pool their insights. This article touches four perspectives on broken symmetry and strong correlation in many-electron systems: ground-state and time-dependent density functional, wavefunction, and model Hamiltonian. For brevity, the important Green's function and dynamical mean field perspectives are not discussed here. Artwork by Liyu Ye, with her permission.

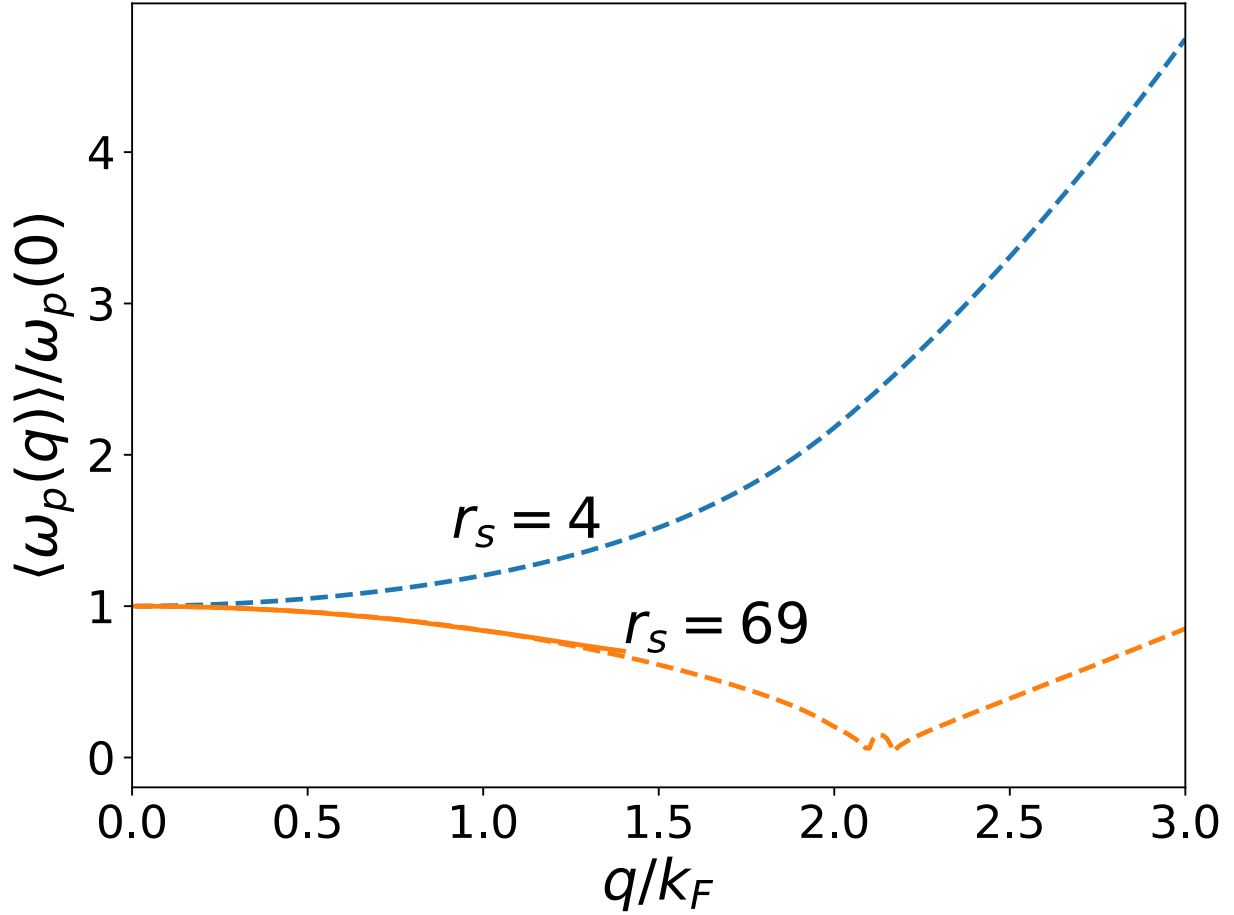

**Fig. 2.** Charge density wave in jellium as a soft plasmon. The frequency $<\omega_p(q)>$ of a density fluctuation as a function of its wavevector $q = 2\pi/wavelength$ in a uniform electron gas at its equilibrium density ($n = 3/[4\pi r_s^3] = k_F^3/(3\pi^2)$; $r_s = 4$) and at a much lower density ($r_s$=69) where a static charge density wave appears and breaks translational symmetry. At $r_s = 4$, the frequency increases with $q$, but at $r_s = 69$ the frequency softens, dropping to zero around $q = 2.28 k_F$, the critical wavevector of the static charge-density wave. The solid part of the $r_s = 69$ curve was computed as in Ref. [27], at wavevectors where the plasmon has not yet penetrated into the continuum of single particle-hole excitations. The dashed parts of the curves were computed here at all wavevectors using the frequency distribution (structure factor [20,21] or spectral function [22,34]) $S(q,\omega) = [-1/(\pi n)] Im \chi_1(q,\omega) > 0$; $\chi_1$ is the density response function at full coupling strength defined in Ref. [27]. The dashes show the average frequency $<\omega_p(q)> = \int_0^\infty \omega S(q,\omega)/\int_0^\infty S(q,\omega)$. The bulk plasma frequency is $\omega_p(0) = [4\pi n]^{1/2}$. For a plot of $S(q) = \int_0^\infty S(q,\omega)$, refer to Fig. S6.

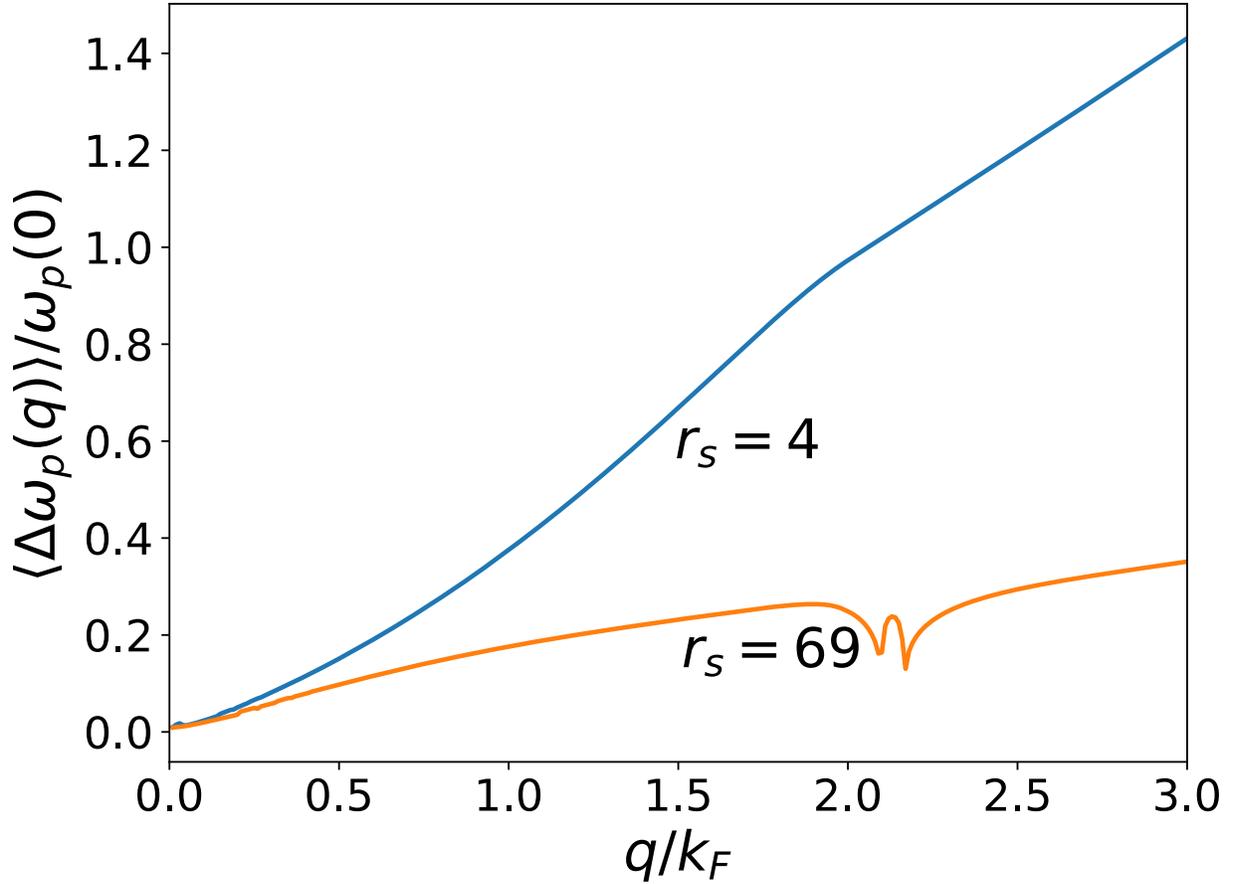

**Fig. 3**. An analogous plot to Fig. 2 presenting the standard deviation in the frequency of a density fluctuation, with the variance defined as $<\Delta\omega_p(q)>^2 = \int_0^\infty \omega^2 S(q,\omega)d\omega / \int_0^\infty S(q,\omega)d\omega - \langle\omega_p(q)\rangle^2$, as a function of its wavevector $q$. Some noise is present at small $q$, where the spectral function is small for all values of $\omega$. At $r_s = 4$, the variance is monotonic, but at $r_s = 69$, the variance begins decreasing as $q$ approaches its critical value ($2.28k_F$) for the static charge-density wave. There is a rather sudden increase in the width of the spectral function as the plasmon enters the continuum of single particle-hole excitations (at $q/k_F > 0.9$ and $1.4$ for $r_s = 4$ and $69$ respectively) that is not reflected in the variance. The variance is controlled by a much higher and narrower central peak. See the contour plots of $S(q,\omega)$ (and the dielectric function) in the Supplementary Materials.

**Table 1.** Mean absolute errors (MAE) in electron volts for the atomization energies of the six representative AE6 [3] *sp*-bonded molecules, for Hartree-Fock exchange (x) and for exchange-correlation (xc) functionals on the first three rungs of a ladder of approximations (none of them fitted to bonded systems). Hartree-Fock results from Lynch and Truhlar [3]. Local spin density approximation (LSDA) [1,4]*,* Perdew-Burke-Ernzerhof (PBE) [5] generalized gradient approximation (GGA), and strongly constrained and appropriately normed (SCAN) [6] meta-GGA results from Bhattarai et al. [7]. Note that the semilocal approximations LSDA, PBE, and SCAN are much better here for xc together than for x or c separately, due to an understood cancellation between the full nonlocalities of exact x and exact c. The LSDA exchange-correlation energy density depends only on the local spin densities, the GGA further includes the density gradients, and the meta-GGA still further includes the orbital kinetic energy densities.

| Approximation | AE6 MAE (eV) |
| --- | --- |
| Hartree-Fock x | 6.3 |
| LSDA xc | 3.3 |
| PBE GGA xc | 0.6 |
| SCAN meta-GGA xc | 0.1 |

# Supplementary Materials for:
# Interpretations of ground-state symmetry breaking and strong correlation in wavefunction and density functional theories

John P. Perdew, Adrienn Ruzsinszky, Jianwei Sun, Niraj K. Nepal, and Aaron D. Kaplan

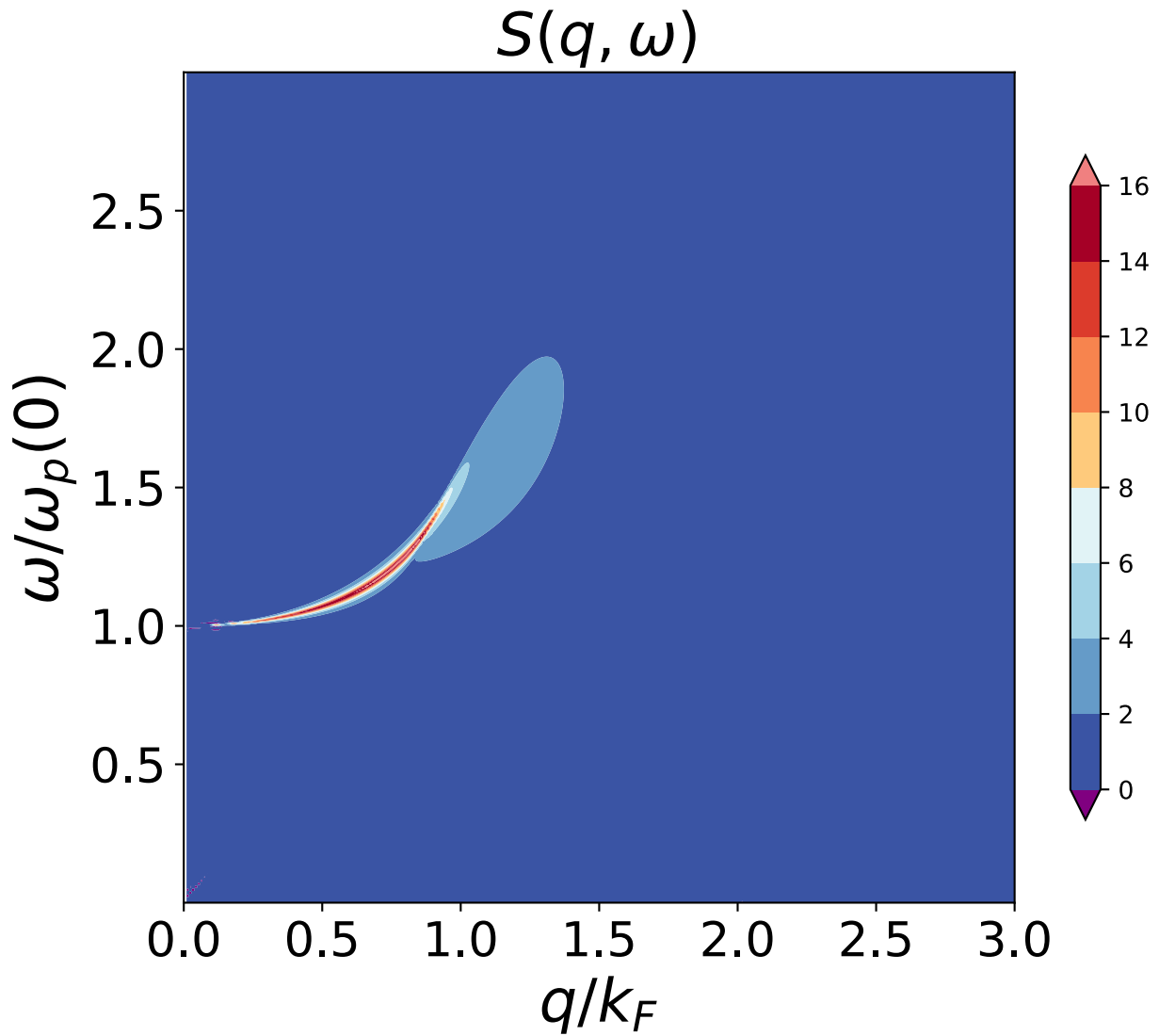

**Fig. S1.**
Spectral function or structure factor $S(q, \omega)$ as defined in the main text for a jellium of bulk density parameter $r_s = 4$.

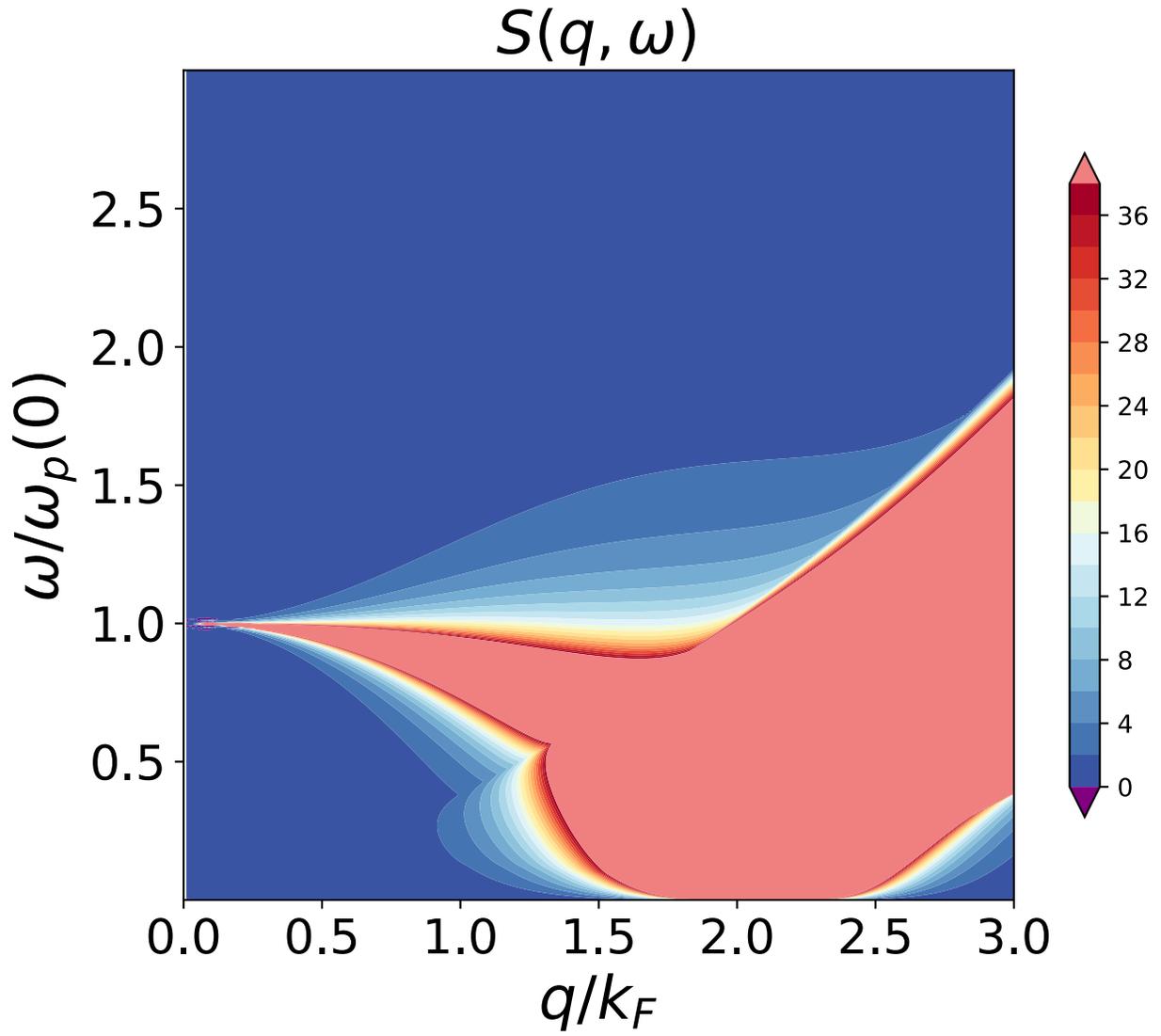

**Fig. S2.**
Spectral function or structure factor $S(q,\omega)$ as defined in the main text for a jellium of bulk density parameter $r_s = 69$.

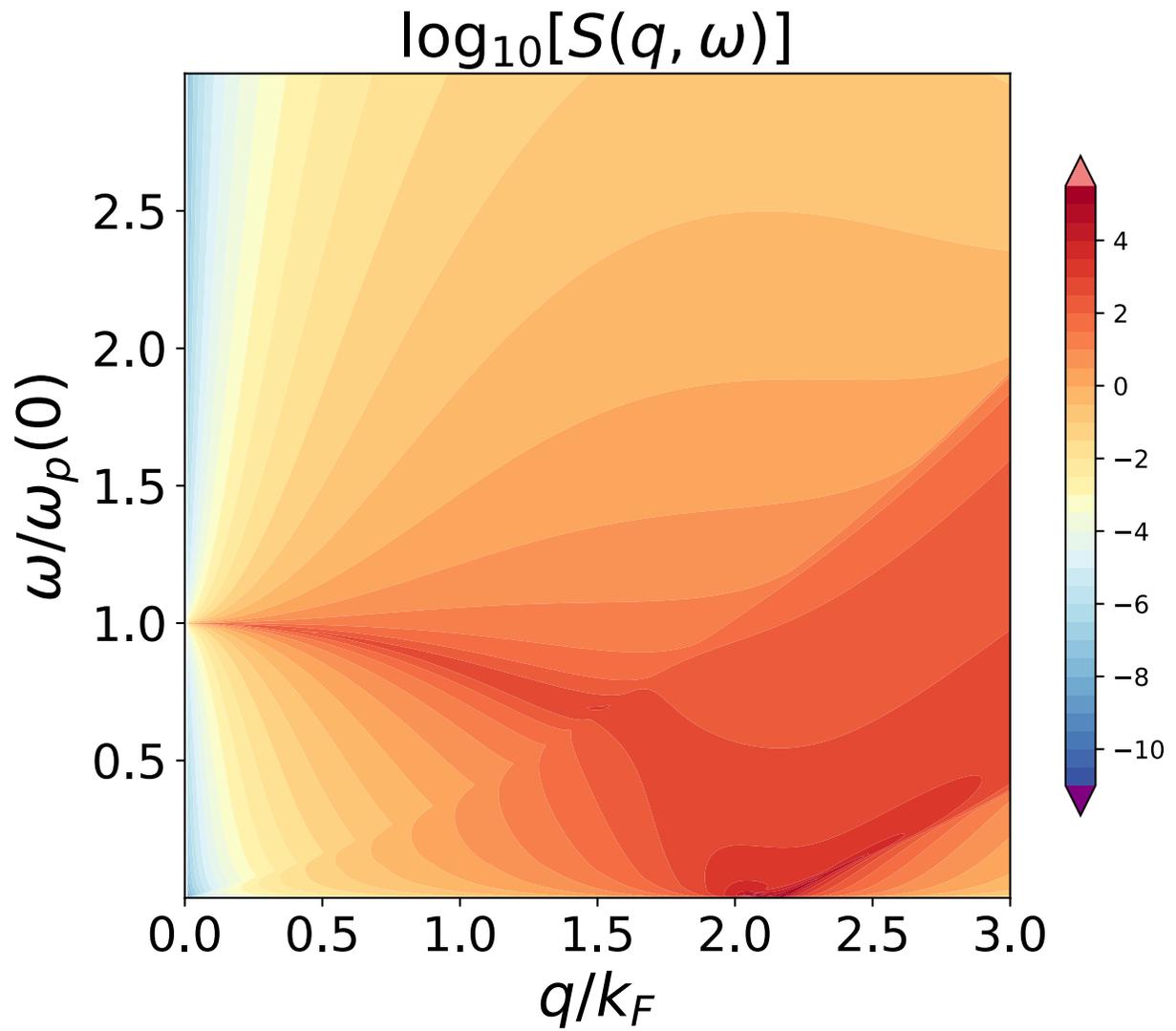

**Fig. S3.**
Same as Fig. S2 but plotting the logarithm of the spectral function.

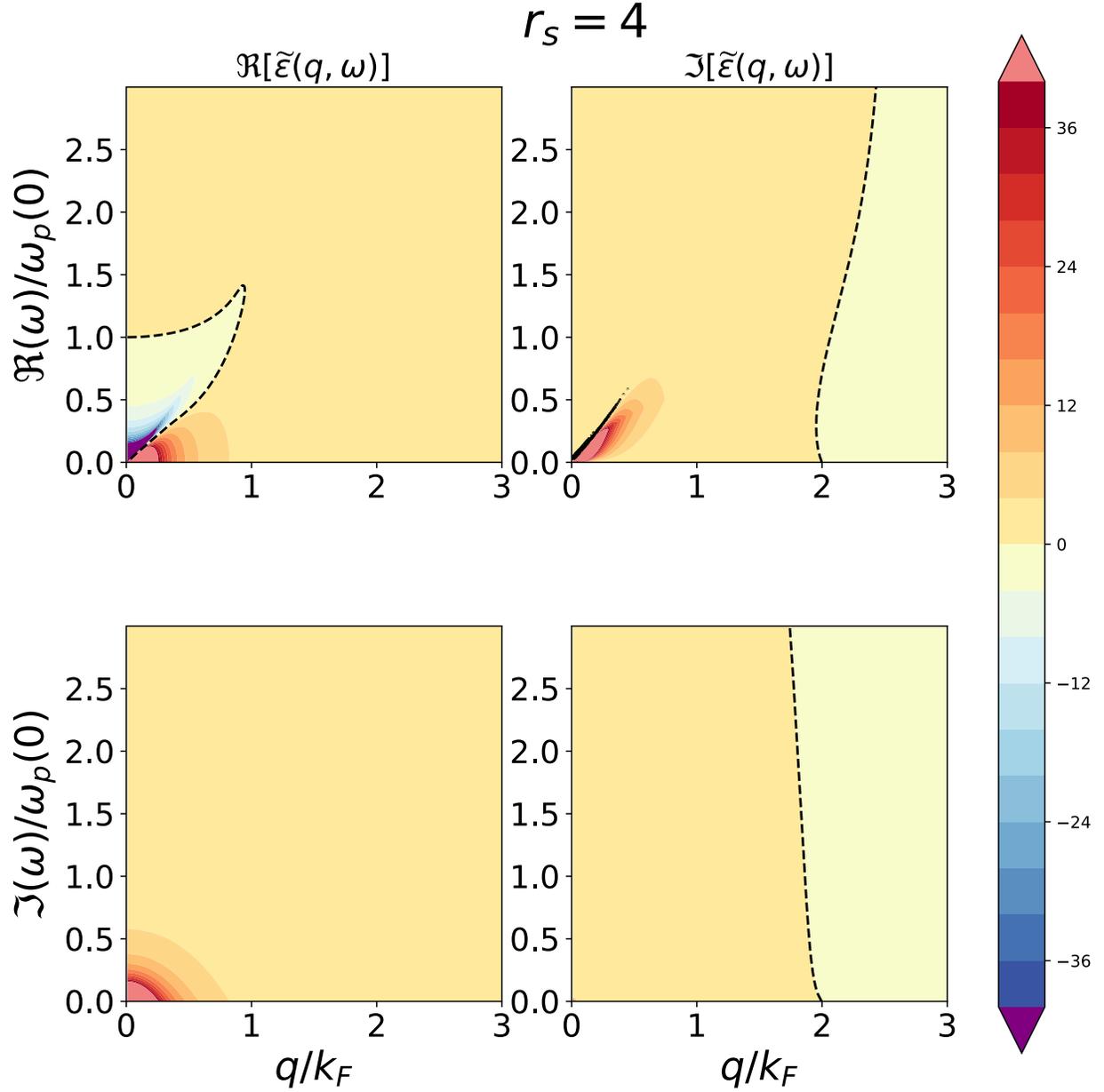

**Fig. S4.**

Real (left) and imaginary (right) parts of the dielectric function $\tilde{\epsilon}(q,\omega) = 1 - \left[\frac{4\pi}{q^2} + f_{xc}(q,\omega)\right]\chi_{KS}(q,\omega)$ in a jellium of $r_s = 4$. $f_{xc}(q,\omega)$ is the exchange-correlation kernel of the MCP07 functional, and $\chi_{KS}(q,\omega)$ is the Lindhard susceptibility [27]. Note that the interacting response function satisfies $\chi(q,\omega) = \chi_{KS}(q,\omega)/\tilde{\epsilon}(q,\omega)$. In the upper panels, $\Im(\omega) = \omega_p(0) \times 10^{-4}$, and in the bottom panels, $\Re(\omega) = \omega_p(0) \times 10^{-4}$. The real or imaginary parts of $\tilde{\epsilon}(q,\omega)$ vanish on the black dashed curves. The upper black dashed curve in the top left panel resembles the real part of the complex plasmon frequency $\omega(q)$ defined by $\tilde{\epsilon}(q,\omega) = 0$ for complex $\omega$ in Ref. [27].

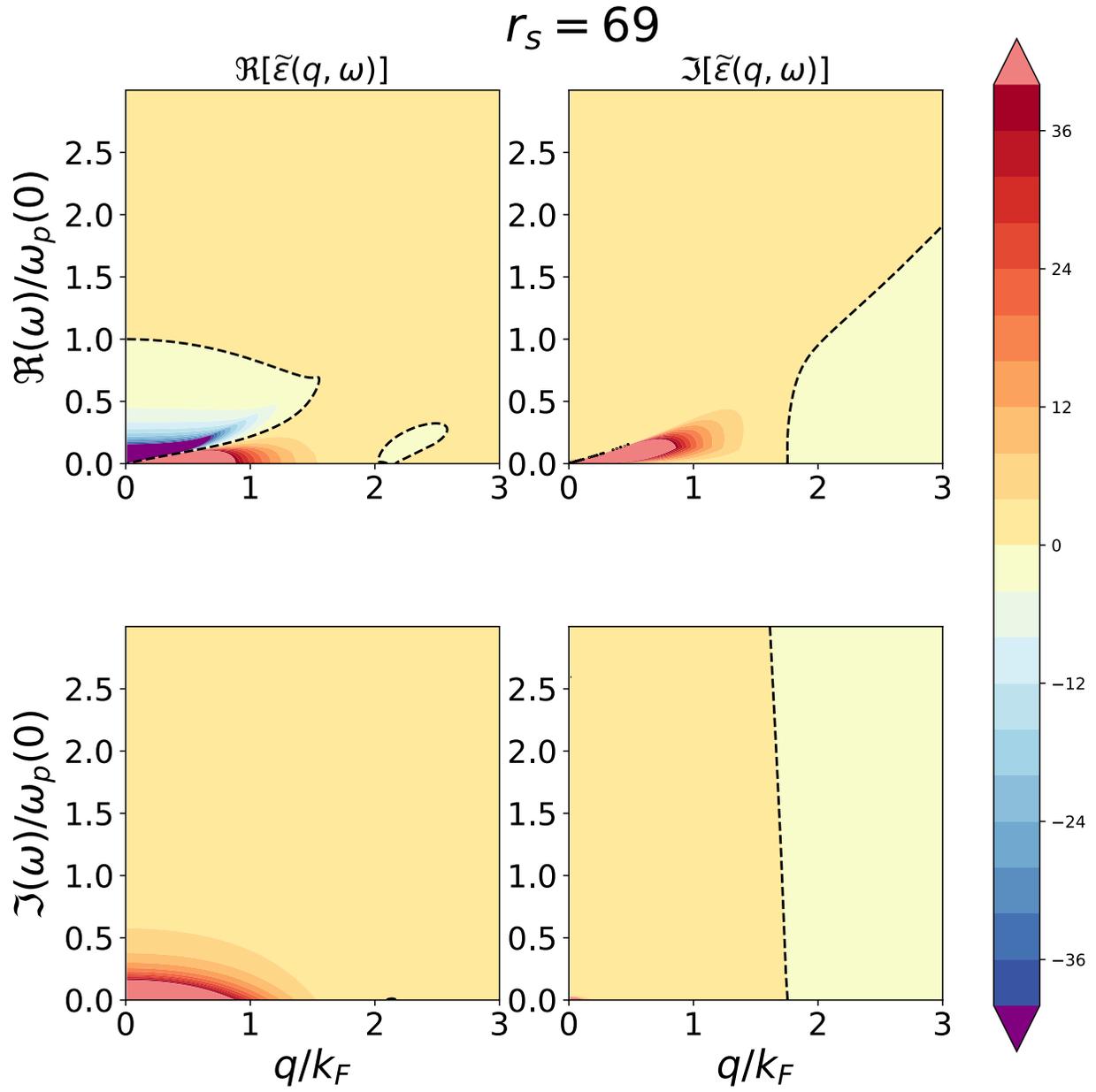

**Fig. S5.**
Same as Fig. S4, but with $r_s = 69$.

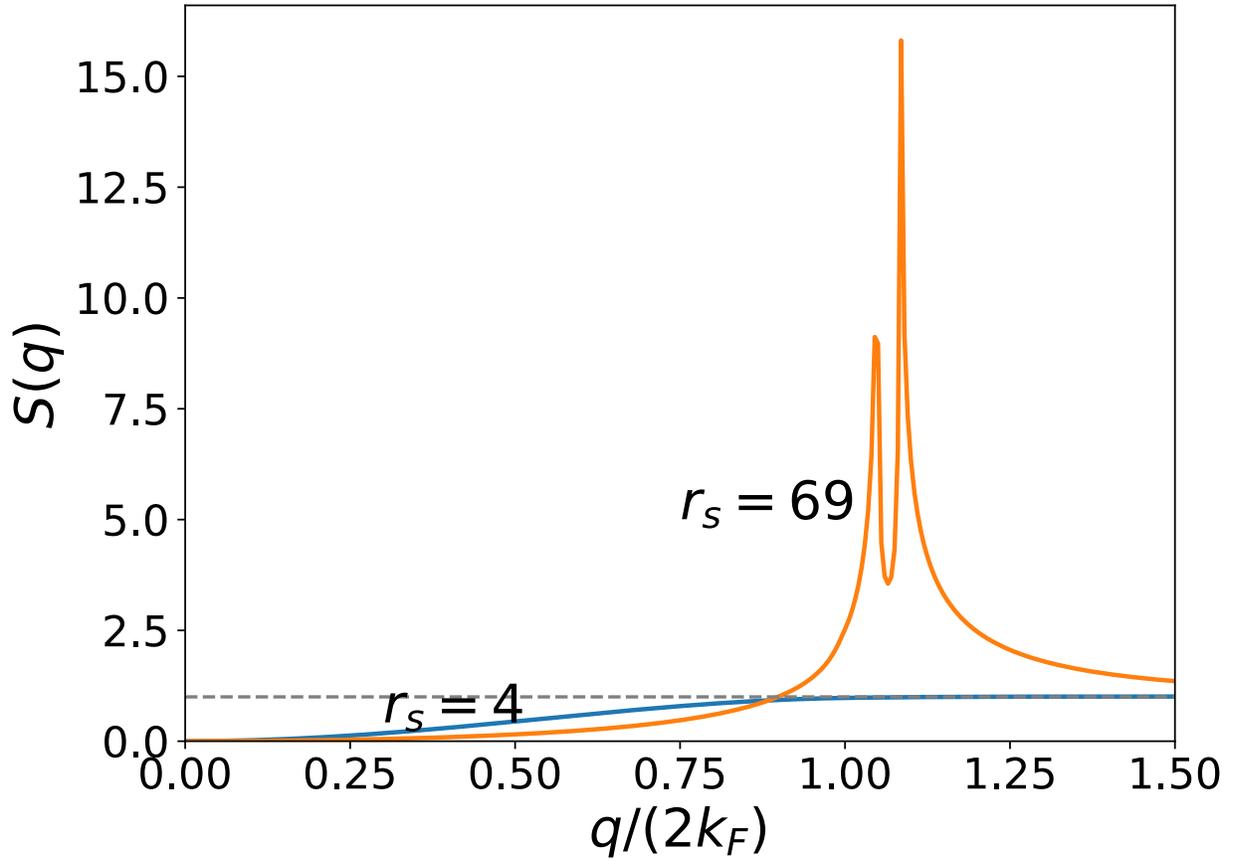

**Fig. S6.**
The static structure factor, $S(q) = \int_0^\infty S(q,\omega)d\omega$ for jellium, which should vary from 0 at $q \to 0$ to 1 at $q \to \infty$ (the gray dashed line is $S(q) = 1$). At $r_s = 4$ (roughly the valence electron density of solid sodium), $S(q)$ exhibits smooth, slowly-varying behavior in $q$. At $r_s = 69$ (a much lower density), $S(q)$ peaks up sharply around the wavevector of the incipient static charge-density wave. For the relationship between $S(q)$ and the pair density $n^2 g(|\mathbf{r}' - \mathbf{r}|)$, see Eq. (6) of Ref. [35].

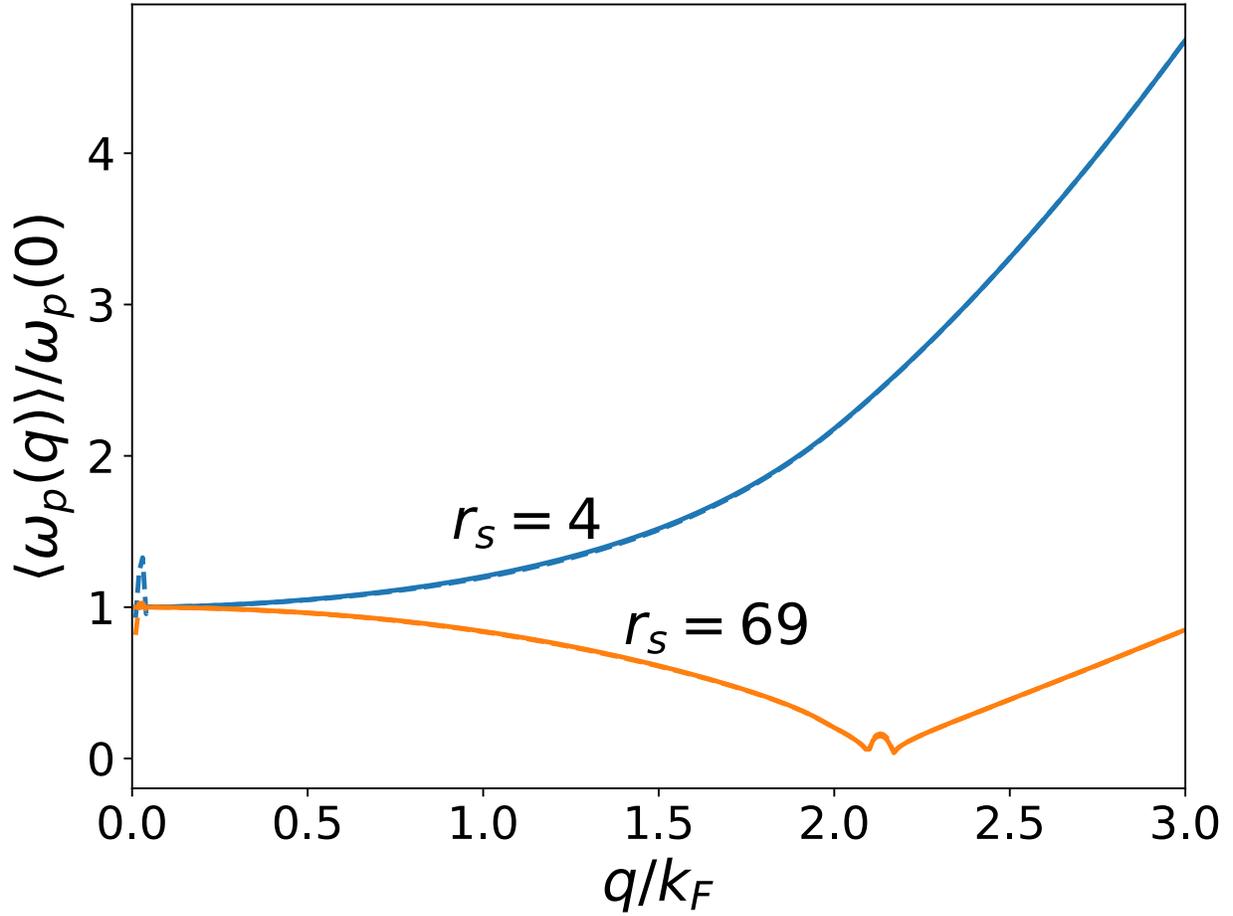

**Fig. S7.**
From the sum rule of Eq. 3.141 of Ref. [22], the exact $<\omega_p(q)>$ as defined in Fig. 2 of the main text equals $q^2/[2S(q)]$. This plot of $q^2/[2S(q)]$ (dashed) and $<\omega_p(q)>$ (solid) demonstrates that the approximate susceptibility used throughout this work [27] quite accurately recovers the sum rule on the exact susceptibility for most values of $q$.